\documentstyle[aps,prl,multicol,psfig,times,fancyheadings]{revtex}
\newcommand{\be}{\begin{equation}}
\newcommand{\ee}{\end{equation}}
\newcommand{\bea}{\begin{eqnarray}}
\newcommand{\eea}{\end{eqnarray}}

\begin{document} 

\title{ Hysteresis in the Random Field Ising Model and Bootstrap
Percolation }

\author{ Sanjib Sabhapandit$^1$ \and Deepak Dhar$^1$ \and Prabodh
Shukla$^2$ } \address{$^1$Department of Theoretical Physics, Tata
Institute of Fundamental Research, Mumbai-400005,
India.  }  \address{$^2$Physics Department, North Eastern Hill
University, Shillong-793 022, India.  }

\maketitle

\begin{abstract} 

We study hysteresis in the random-field Ising model with an asymmetric
distribution of quenched fields, in the limit of low disorder in two
and three dimensions. We relate the spin flip process to bootstrap
percolation, and show that the characteristic length for
self-averaging $L^{\star}$ increases as $\exp(\exp (J/\Delta))$ in 2d,
and as $\exp(\exp(\exp(J/\Delta)))$ in 3d, for disorder strength
$\Delta$ much less than the exchange coupling $J$. For system size $1
\ll L < L^{\star}$, the coercive field $h_{_{coer}}$ varies as $2 J -
\Delta \ln \ln L$ for the square lattice, and as $ 2J - \Delta \ln \ln
\ln L$ on the cubic lattice.  Its limiting value is $0$ for $L
\rightarrow \infty$ both for square and cubic lattices. For lattices
with coordination number $3$, the limiting magnetization shows no
jump, and $h_{_{coer}}$ tends to $J$.

\vskip3mm
\noindent 
DOI:10.1103/PhysRevLett.88.197202  \hfill 

\end{abstract}

\setlength{\headrulewidth}{1pt}
\lhead{Physical Review Letters {\bf 88}, 197202 (2002)}
\rhead{\thepage}
\cfoot{}
\thispagestyle{fancy}
\pagestyle{fancy}

\begin{multicols}{2} 

In recent years, there has been a lot of interest in the study of
hysteresis in magnetic systems, both theoretically~\cite{theory}, and in
experiments~\cite{experiments}. Hysteresis in the random-field Ising model
(RFIM) model was first discussed by Sethna {\it et al}~\cite{SETHNA}, who
proposed it as a model of return point memory, and of Barkhausen
noise~\cite{barkhausen}. Sethna {\it et al} solved the model in the
mean-field limit, and showed that if the strength $\Delta$ of the quenched
random field is large, the average magnetization per site is a continuous
function of the external field, but for small $\Delta$, it shows a
discontinuous jump as the external field is increased.  Interestingly, the
non-equilibrium hysteresis response in the RFIM can be determined exactly
on a Bethe lattice~\cite{SHUKLA-1D,DHAR}, though the corresponding
equilibrium problem has not been solved so far, even in zero field. These
calculations have been extended to determining the distribution of sizes
of the Barkhausen jumps~\cite{SABHAPANDIT}, and the calculation of minor
hysteresis loops~\cite{minorloops,DANTE}.

In this paper, we study the low disorder limit of the hysteresis loop
in the RFIM on periodic lattices in two and three dimensions.  We find
that the behavior of hysteresis loops depends nontrivially on the
coordination number $z$. For $z=3$, for continuous unbounded
distributions of random fields, the hysteresis loops show no jump
discontinuity of magnetization even in the limit of small disorder,
but for higher $z$ they do. This is exactly as found in the exact
solution on the Bethe lattice \cite{DHAR}.

The analytical treatment of self-consistent equations on the Bethe
lattice is immediately generalized to asymmetrical case. However, we
find that behavior of hysteresis loops in euclidean lattices can be
quite different from that on the Bethe lattice, for asymmetrical
distributions. On hypercubical lattices in $d$ dimensions, there is an
instability related to bootstrap percolation, that is absent on the
Bethe lattice. This reduces the value of the coercive field
$h_{_{coer}}$ away from the Bethe lattice value ${\cal O}(J)$ to zero,
where $J$ is the exchange coupling.  We note that the limit $\Delta
\rightarrow 0$ is somewhat subtle, as the system size $L^{\star}$
required for self-averaging diverges very fast for small $\Delta$, and
the finite-size corrections to the thermodynamic limit tend to zero
very slowly.

In the RFIM, the Ising spins $\{s_i\}$ with nearest neighbor
ferromagnetic interaction $J$ are coupled to the on-site quenched
random magnetic field $h_i$ and the external field $h$. The
Hamiltonian of the system is given by
\be H=-J 
\sum_{<i,j>} s_is_j -\sum_{i}h_is_i -h\sum_{i}s_i .
\label{HAMILTONIAN}
\ee
We assume that $\{h_i\}$ are quenched independent identically
distributed random variables with the probability that the value of
the random field at site $i$ lies between $h_i$ and $h_i + dh_i$ being
$\phi(h_i)\, dh_i$.

The system evolves under the zero-temperature Glauber single-spin-flip
dynamics~\cite{KAWASAKI}: a spin-flip is allowed only if the process
lowers energy.  We assume that the rate of spin-flips is much larger
than the rate at which $h$ is changed, so that all flippable spins may
be said to relax instantly, and any spin $s_i$ always remains parallel
to the net local field $\ell_{i}$ at the site:
\be
 s_i= \mbox{sign}(\ell_{i}) = \mbox{sign}( J \sum_{j=1}^{z} s_{j} + h_{i}
+ h) .
\label{LOCAL_FIELD} 
\ee 

Under this dynamics, for ferromagnetic coupling ($J > 0$), if we start
with any stable configuration, and then increase the external field
and allow the system to relax, the final stable configuration reached
is  independent of the order in which the unstable spins are
flipped.  Also, in the relaxation process no spin flips more than
once.

For a given distribution $\phi(h_i)$, we define $p_m(h)$ with $0 \leq m
\leq z$ as the conditional probability that the local field at any site
$i$ will be large enough so that it will flip up, if $m$ of its neighbors
are up, when the uniform external field is $h$.  Clearly
\be
p_{m}(h)=\int_{(z-2m)J-h}^{\infty} \phi(h_{i}) \, \, dh_{i} .
\label{p_m}
\ee
Clearly, for any given value of $h$, the magnetization depends
on the distribution $\phi(h_i)$ only through $p_m(h)$.

Historically, RFIM was first studied in the context of possible
destruction of long range order by arbitrarily weak quenched disorder
in equilibrium systems. Accordingly the distribution of random field
was assumed to be symmetrical. However, in hysteresis problem, the
symmetry between up and down spins state is already broken by the
specially prepared initial state ( all down in our case), and the
symmetry of the distribution plays no special role.  In the following,
we shall assume that the distribution has a asymmetrical shape, given
by
\be
\phi(h_i) = \frac{1}{\Delta}  \exp(- h_i/\Delta) \theta(h_i);
\label{RF_dist}
\ee
where $\theta$ is the step function. The mean value of $h_i$ can be
made zero by a shift in the value of the external uniform field.  Our
treatment is easily extended to other continuous unimodal
distributions. The exact form of $\phi(x)$ is not important, and other
forms like $\exp(-x -e^{-x})$ which fall sharply for negative $x$ have
the same behavior.


Consider first the case of the two-dimensional hexagonal lattice with
$z =3$. For periodic boundary conditions (PBC), if $\Delta=0$,
starting with a configuration with all spins down, clearly one has
$h_{_{coer}} =3J$. For $\Delta \neq 0$, the site with the largest
local field flips first, and then if $h > J$, $p_1(h) = 1$, this
causes neighbors of the flipped spin to flip, and their neighbors, and
so on. Thus, so long as there is at least one flipped spin, all other
spins also flip, and the magnetization is $1$. The largest local field
in a system of $L^2$ spins is of order $ 2 \Delta \ln L$. Once this
spin turns up, other spin will flip also up, causing a jump in
magnetization from a value $\approx -1$ to a value $ +1$ in each
sample. Hence the coercive field, (the value of $h$ where
magnetization changes sign) {\it to lowest order in $\Delta$}, is
given by
\be
h_{_{coer}} = 3 J - 2 \Delta \ln L, ~~~\mbox{for}~~ 1
\ll \ln L \ll J/\Delta.
\ee
Sample to sample fluctuations in the position of the jump are of order
$\Delta$. On averaging over disorder, the magnetization will become a
smooth function of $h$, with the width of the transition region being
of order $\Delta$.

For a fixed $\Delta \ll J$, if $L$ is increased to a value near $
\exp(J/\Delta) \equiv L^{\star}_{hex}$, $h_{_{coer}}$ decreases to a
value near $J$. For $h \approx J$, $p_1(h)$ is no longer nearly $1$,
but $p_0(h) \simeq 0, p_2(h) \simeq p_3(h) \simeq 1$. The value of
magnetization depends only on $p_1(h)$, which is a function of
$\widetilde{h} = ( h - J)/\Delta$. As $\widetilde{h}$ increased from
$-\infty$, $p_1(h)$ increases continuously from $0$ to $1$.

\begin{figure}
\begin{center}
\leavevmode
\psfig{figure=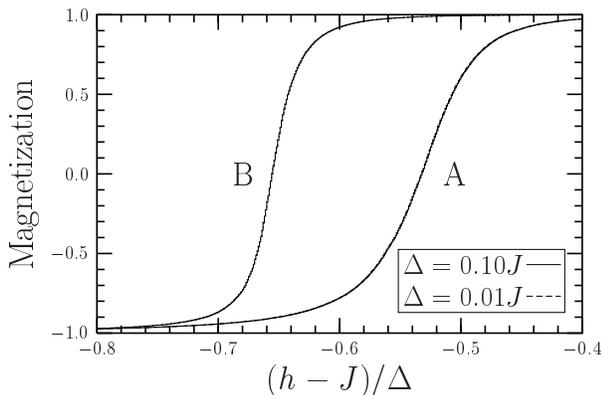,width=8cm,angle=0}
\caption{\small Magnetization in the increasing field.  The curves for
the two values of $\Delta$ coincide. Curves A and B are for 2-d and
3-d lattice with $z=3$.}
\label{HEXAGONAL_MAGNETIZATION}
\end{center}
\end{figure}

In Fig.~\ref{HEXAGONAL_MAGNETIZATION}, curve A shows the result of a
simulation on the hexagonal lattice with $L=4096$, and PBC.
To avoid the problem of probability of nucleation
being very small for $h$ near $J$, we made the local field at a small
fraction of randomly chosen sites very large, so that these spins are
up at any $h$.  The number of such spins is of order $L$, so that
their effect on the average magnetization is negligible. Introduction
of these ``nucleation centers" makes $L^{\star} \approx {\cal {O}}
(\sqrt{L})$ ( the average separation between centers), and
$h_{_{coer}}$ drops to a value near $J$, so that, we can study the
large $L$ limit with available computers.  For $ L > L^{\star}_{hex}$,
the behavior of hysteresis loops becomes independent of $L$.

We see that magnetization no longer undergoes a single large jump, but
many small jumps. In the figure, we also show the plot of
magnetization when the random field at each site is decreased by a
factor 10. This changes the value $\Delta$ from $0.1J$ to
$0.01J$. However, plotted as a function of $\widetilde{h}$, the
magnetization for these two different values (for small $\Delta$) fall
on top of each other {\em for the same realization of disorder}
(except for the overall scale $\Delta$).  Thus we can decrease
$\Delta$ further to arbitrarily small values, and the limit of $\Delta
\rightarrow 0$ is straightforward for each realization of
disorder. Then, averaging over disorder, for a fixed $\Delta$, we see
that $h_{_{coer}}$ tends to the value $J$ as $\Delta$ tends to $0$.
Also, we see that there is no macroscopic jump-discontinuity for any
non-zero $\Delta$.

We also show in Fig.~\ref{HEXAGONAL_MAGNETIZATION} [curve B], the
results of simulation of a $3$-dimensional lattice with $z=3$ of size
$256^3$ with PBC. The behavior is qualitatively same as that in two
dimensions. The value of $h_{_{coer}}=J$ in the limit $\Delta
\rightarrow 0$ is same for symmetrical distribution, and also is the
same as predicted by the Bethe approximation.

\begin{figure}
\begin{center}
\leavevmode
\psfig{figure=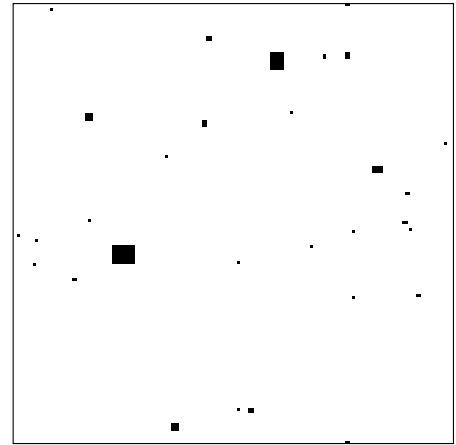,width=6cm,angle=0}
\caption{\small A snapshot of the up-spins just before the jump
($h=1.998243J$). The lattice size is $200\times 200$ and
$\Delta=0.001J$. Initial configuration is prepared with $0.05\%$
up-spins.}
\label{CONFIGURATION}
\end{center}
\end{figure}

\begin{figure}
\begin{center}
\leavevmode
\psfig{figure=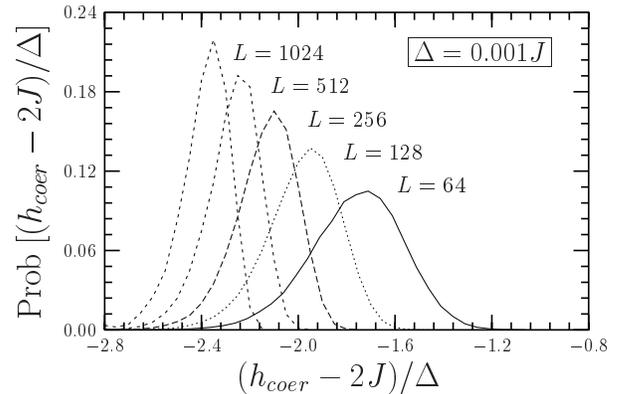,width=8cm,angle=0}
\caption{\small Distribution of the scaled coercive field on a square
lattice for different lattice size $L^2$.}
\label{frequency}
\end{center}
\end{figure}

On the square lattice also, the value of $h_{_{coer}}$ is determined
by the need to create a nucleation event. Arguing as before, we see
that $h_{_{coer}}$ to lowest order in $\Delta$ is given by $
h_{_{coer}} \approx 4 J - 2 \Delta \ln L$, for $1 \ll \ln L \ll
J/\Delta$. Adding a small number of nucleation sites suppresses this
slow transient, and lowers $h_{_{coer}}$ from $4 J$ to a value near $2
J$.  However, in this case, even after adding the nucleation centers,
the system shows a large single jump in magnetization, indicating the
existence of another instability.  We observed in the simulation that
at low $\Delta$, as $h$ is increased, the domains of up spins grow in
rectangular clusters [see Fig.~\ref{CONFIGURATION}] and at a critical
value of $h_{_{coer}}$, one of them suddenly fills the entire lattice.
This value $h_{_{coer}}$ fluctuates a bit from sample to sample.  In
Fig.~\ref{frequency} we have plotted the distribution of the scaled
variable $\widetilde{h_c} = (h_{_{coer}}-2J)/\Delta$ for different
system sizes $L$, for $\Delta=0.001 J$. The number of different
realizations varies from $10^4$ (for the largest $L$) to $10^5$ (for
the smallest $L$).  Note that the distribution shifts to the left with
the increasing system size, and becomes narrower.

This instability can be understood as follows: on a square lattice,
for the asymmetric distribution [Eq.~\ref{RF_dist}] for $h >0 $, $p_m
=1$ for $m\ge 2$, and any spins with more than one up-neighbors flips
up. Therefore, stable clusters of up spins are rectangular in shape.
The growth of domains of up spins is same as in the bootstrap
percolation process BP$_m$ with $ m = 2$
~\cite{CHALUPA,AIZENMAN,ADLER}.  In the process BP$_m$, the initial
configuration is prepared by occupying lattice sites independently
with a probability $p$ and the resulting configuration is evolved by
the rules: the occupied sites remain occupied forever, while an
unoccupied site having at least $m$ occupied neighbors, becomes
occupied. For $m=2$, on a square lattice, in the final configuration,
the sites which are occupied form disjoint rectangles, like the
cluster of up-spins in Fig.~\ref{CONFIGURATION}. It has been proved
that in the thermodynamic limit of large $L$, for any initial
concentration $p>0$, in the final configuration all sites are occupied
with probability $1$~\cite{AIZENMAN}.

Now consider a rectangular cluster of up spins, of length $l$ and
width $m$.  Let $P(l,m)$ be the probability that, if this
rectangle is put in a randomly prepared background of density $p_1(h)$,
this rectangle will grow by the BP$_2$ process to fill the entire
space. The probability that the random fields at any sites neighboring
this rectangle will be large enough to cause it to flip up is
$p_1(h)$. The probability that there is at least one such site along
each of two adjacent sides of length $l$ and $m$ of the rectangle is
$(1-q^l)(1-q^m)$, where $q = 1 - p_1(h)$. Once these spins flip up, this
induces all the other spins along the boundary side to flip up and the
size of the rectangle grows to $(l+1)\times(m+1)$. Therefore
\be
P(l,m) \ge (1-q^l)(1-q^m) P(l+1,m+1).
\label{P(l,m)}
\ee
Thus the probability of occurrence of a nucleation
which finally grows to fill the entire lattice is
\be
P_{nuc} \ge p_0(h)\prod_{j=1}^{\infty}(1-q^j)^2 .
\label{P_nuc}
\ee The right hand side can be shown to vary as $
p_0(h)\exp\left(-{\pi^2 \over 3 p_1(h)}\right)$ for small $p_1(h)$.
The condition to determine $h_{_{coer}}$ is that for this value of
$h$, $P_{nuc}$ becomes of order $1/L^2$, so that we get
\be
p_0(h_{_{coer}}) \exp\left(-{\pi^2 \over 3 p_1(h_{_{coer}})}\right)
\approx {1 \over L^2}. 
\label{hcoersq}
\ee 
This equation can be solved for $h_{_{coer}}$ for any given $L$. For
the distribution given by Eq.~(\ref{RF_dist}), this becomes
\be
\exp\left(\frac{ h_{_{coer}} -4J}{\Delta}\right) \exp\left[-{2\pi^2 \over 3}
\exp\left(\frac{-h_{_{coer}} +2 J}{\Delta}\right)\right] \approx {1 \over L^2} .
\ee 
It is easy to see from this equation that for  $1~\ll~\ln L~\ll~J/\Delta$,
the leading $L$-dependence of $h_{_{coer}}$, to lowest order in $\Delta$
is given by
\be
h_{_{coer}} \approx 4J - 2\Delta \ln L;
\ee
and for  $J/\Delta \ll \ln L \ll \exp( 2 J/\Delta)$, 
\be
h_{_{coer}} \approx 2 J - \Delta\ln[\frac{3}{\pi^2} (\ln L - J/\Delta)].
\label{h_c}
\ee
This agrees with our observation that the
scaled critical field $\widetilde{h_c}$ shifts to the left with 
increasing system size.

\begin{figure}
\begin{center}
\leavevmode 
\psfig{figure=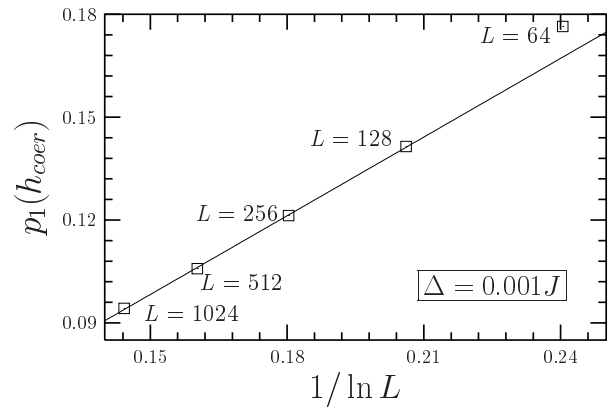,width=8cm,angle=0}
\caption{\small $p_1(h_{_{coer}})$ vs. $1/\ln L$ for square lattice.}
\label{p-logL}
\end{center}
\end{figure}

To test the validity of Eq.~(\ref{hcoersq}) in simulations, 
we put
$p_0(h)= 0.005$ independent of $h$. Eq.~(\ref{hcoersq}) then
simplifies to
\be
p_1(h_{_{coer}}) \approx {\pi^2 \over 6 \ln L} .
\label{p_1(h_c)} 
\ee
In Fig.~\ref{p-logL}, we have plotted $p_1$  for the mean
$h_{_{coer}}$ from Fig.~\ref{frequency} versus $1/\ln L$. The graph is
approximately a straight line, which agrees with Eq.~(\ref{p_1(h_c)}).
The slope of the line is less than in Eq.~(\ref{p_1(h_c)}), which only
gives an upper bound to  $h_{_{coer}}$.

If $h>0$, we will have $p_2=1$, and bootstrapping ensures that so long
as $p_0>0$, we will have all spins up in the limit of large $L$. This
implies that  $h_{_{coer}}=0$ in this limit.

If there are sites with large negative quenched fields, the bootstrap
growth stops at such sites. Hence the bootstrap instability cannot be
seen for symmetric distributions. Even if the quenched fields are only
positive, the instability does not occur on lattices with $z = 3$. On
such lattices, if the unoccupied sites percolate, there are infinitely
extended lines of unoccupied sites in the lattice. These cannot not
become occupied by bootstrapping under BP$_2$.  Thus the critical
threshold for BP$_{2}$ on such lattices is not $0$.


The above analysis is easily extended to higher dimensions. In $d=3$,
if $h >0$, then $p_m(h) = 1$ for $m\ge 3$, therefore the spin flip
process is similar to the spanning process of three dimensional
BP$_3$~\cite{CERF}.  But in this case, it is known that for any
initial non-zero density, in the thermodynamical limit, the final
configuration has all sites occupied with probability $1$.  The
clusters of up-spins grow as cuboids, and at each surface of the
cluster, the nucleation process is similar to that in two dimension.
Let $\epsilon$ be the probability that, a nucleation occurs at a given
point of a surface of the clusters of up spins which sweeps the entire
two dimensional plane at $h$.
\be 
\epsilon \approx p_1(h) \exp\left(-{\pi^2 \over 3 p_2(h)}\right) . 
\ee
The probability that, there exist at least one nucleation which sweeps
the entire plane of size $l\times l$, is
$1-(1-\epsilon)^{l^2}$. Therefore, the probability $P_{nuc}$, that a
nucleation sweeps the entire three dimensional lattice at $h$
satisfies
\be
P_{nuc} \ge
p_0(h)\prod_{l=1}^{\infty}\left[1-(1-\epsilon)^{l^2}\right]^3 .
\ee
For small $\epsilon$, the infinite product can be shown to vary as
$\exp(-A/ \sqrt{\epsilon})$, with $A ={3\over2}\sqrt{\pi}\zeta(3/2)$.

 $h_{_{coer}}$ is determined by the condition that $P_{nuc}$
must be of the
order $1/L^3$:
\be
p_0(h_{_{coer}}) \exp\left[- \frac{A}{\sqrt{p_1(h_{_{coer}})}} \exp\left(
\frac{\pi^2}{6 p_2(h_{_{coer}})}\right)\right] \approx 1/L^3 .
\ee

The leading L-dependence of $h_{_{coer}}$ is different in different
ranges of $h_{_{coer}}$, depending on whether the strongest dependence
of the left-hand side comes from variation of $p_0(h), p_1(h)$ or
$p_2(h)$.  We find that, $h_{_{coer}} \approx 6J - 3 \Delta \ln L$,
for $4 J < h_{_{coer}} <6 J$. It is $\approx 4J - 2 \Delta \ln (\ln L
- {2 J/3\Delta})$, for $2 J < h_{_{coer}} < 4 J$; and $\approx 2J -
\Delta \ln \ln (\ln L -{2J/3 \Delta})$, for $\Delta \ll h_{_{coer}} <
2 J$.  It is straightforward to determine the corresponding ranges of
$L$ for the validity of these equations.

In the limit $L \gg L^{\star}_{cub} = \exp( \exp( \exp( 2
J/\Delta)))$, the loop becomes independent of $L$, with $h_{_{coer}}
\rightarrow 0$. We have also verified the existence of jump in
numerical simulation for $z=4$ (diamond lattice) in three dimensions.

In brief, we have shown that the hysteresis loops on lattices with
coordination number three are qualitatively different from those with
$z>3$. For the square and cubic lattices, $h_{_{coer}}$ decreases to
$\Delta$ very slowly for large $L$. In general, it is true for
lattices where the corresponding bootstrap percolation problem has an
instability.

We thank M. Barma and N. Trivedi for critically reading  the
manuscript.


\end{multicols}
\end{document}